\documentclass[12pt]{article}
\usepackage{amssymb,amsmath,epsfig}
 \allowdisplaybreaks

\begin{document}
\title{\bf Causes of Irregular Energy Density in $f(R,T)$ Gravity}

\author{Z. Yousaf$^1$ \thanks{zeeshan.math@pu.edu.pk}, Kazuharu Bamba$^2$
\thanks{bamba@sss.fukushima-u.ac.jp} and M. Zaeem ul Haq
Bhatti$^1$ \thanks{mzaeem.math@pu.edu.pk}\\
$^1$ Department of Mathematics, University of the Punjab,\\
Quaid-i-Azam Campus, Lahore-54590, Pakistan\\
$^2$ Division of Human Support System,\\ Faculty of Symbiotic Systems Science,\\ Fukushima University, Fukushima 960-1296, Japan}

\date{}

\maketitle
\begin{abstract}
We investigate irregularity factors for a self-gravitating spherical
star evolving in the presence of imperfect fluid. We explore the
gravitational field equations and the dynamical equations with the
systematic construction in $f(R,T)$ gravity, where $T$ is the trace
of the energy-momentum tensor. Furthermore, we analyze two
well-known differential equations (which occupy principal importance
in the exploration of causes of energy density inhomogeneities) with
the help of the Weyl tensor and the conservation laws. The
irregularity factors for a spherical star are examined for
particular cases of dust, isotropic and anisotropic fluids in
dissipative and non-dissipative regimes in the framework of $f(R,T)$
gravity. It is found that as the complexity in the matter with the
anisotropic stresses increases, the inhomogeneity factor has more
correspondences to one of the structure scalars.
\end{abstract}
{\bf Report number:} FU-PCG-12

\section{Introduction}

The influence of modification in gravity theories have gained
significant attention due to the motivation in both high energy
physics and cosmology. Since there exist observational evidences
of the accelerating universe \cite{SN, CMB, LSS, Eisenstein:2005su, Jain:2003tba} but still some predictions and
compelling theoretical work about the expansion of the universe is
under consideration. After the successful detection of the
gravitational waves, it is still possible that the cosmological
constant added by Einstein in his field equations can describe the
accelerating phase of the universe. However, the unnatural fine
tuning problem favors the possibility of dark side in the universe
due to the dark dynamical effects. To test the viability of any
gravitational theory, the most significant way is to match their
predictions with real object's motion.

The modification in the Einstein's theory to discuss the dark
effects involves generalization in the Lagrangian of
Einstein-Hilbert action. The simplest generalization is to use
function $f(R)$ instead of Ricci scalar in the action. However, in
order to include the matter contents, a simplest generalization is
to replace $R$ with $f(R,T)$, where $T$ represents the trace of
stress-energy tensor. It is noted that such addition in the
Lagrangian can be observed as the addition of new degrees of
freedom. The equation of motion emerging from such kind of
Lagrangian will differ from the Einstein's one. In that case, it
would be possible to eliminate the cosmological constant to describe
the acceleratory phase of the universe. Such Lagrangians have much
significance to study the dark energy (DE) and dark matter (DM)
problems and much attention has never been made along this
direction (for reviews on the late-time cosmic acceleration, i.e.,
dark energy problem, and modified gravity theories,
see, e.g., \cite{ya3, v41, b2b, b2a, R-DE-MG}).

Theories involving curvature matter coupling have attained
significant attention to explore the enigma of cosmic evolution and
other cosmological aspects. A geometry matter coupled system results
in existence of extra force due to non-geodesic motion of test
particles and such systems in the setting of Lagrangian for $f(R,T)$
have been introduced by \cite{3}. It has been observed that the
extra force vanishes if one used the specific form of the Lagrangian
for usual matter (e.g. $L_m=p$) for non-minimally coupled $f(R)$
theories \cite{nyb4,nyb5}, however, the extra force does not vanish
for matter geometry coupled system. The $f(R,T)$ gravity theory is
considered as a useful candidate to study the acceleratory behavior
during the cosmic expansion which is not only due to the
scalar-curvature part but includes the matter components as well.
This theory is also considered as a useful candidate among the
modified gravities which is based on the non-minimally curvature
matter coupling. This theory came into with the background that the
cosmological constant could be considered as the trace dependent
function, i.e., $\Lambda(T)$ gravity. This is done to make the
interaction between the usual cosmic matter and DE which is
supported by some modern cosmological data \cite{nyb1}.

Faulkner et al. \cite{1} explored the $f(R)$ theory and try to
equate it with the scalar tensor theories with two classes of
models: massive and Chameleon $f(R)$ models. In recent years, it has
been observed that most of the models proposed in $f(R)$ gravity
does not satisfy the weak-field solar system constraints \cite{2}.
Harko et al. \cite{3} extended the Einstein's standard relativity
theory to $f(R,T)$ and investigated that higher-curvature theories
can assists enough to resolve the flatness issue in the rotation
curves of galaxies. The field equations for some specific models
with explicit $f(R,T)$ configuration has also been presented. Reddy
et al. \cite{4} discussed the the Bianchi type III universe model
with perfect matter configuration in the background of $f(R,T)$
gravity to study the early universe. Adhav \cite{5} has found some
interesting Bianchi-I universe model in this theory.

Sharif and Yousaf \cite{nyb7} examined the stability of isotropic
compact objects framed within $f(R,T)$ gravity and found relatively
more stable and compact objects than that observed in $f(R)$
gravity. Sun and Huang \cite{nyb8} analyzed cosmic evolution in
$f(R,T)$ gravity by means of redshift fluctuations against distance
modulus and confronted good fitting numerical plots consistent with
the observational data of astronomy. Baffou et al. \cite{nyb9}
investigated dynamical evolution along with the stability of
power-law and de Sitter cosmic models against linear perturbation.
They concluded that such models can be considered as a competitive
dark energy candidate. Alves et al. \cite{nyb10} explored the
physical behavior of gravitational waves in different formalisms of
$f(R,T)$ gravity models and showed that the gravitational wave
spectrum have strong dependence on $f(R,T)$ model. Alhamzawi and
Alhamzawi \cite{nyb11} explored a considerable contribution of
$f(R,T)$ gravity on gravitational lensing and found comparable
results with those already exists in literature, thereby suggesting
the viability of this theory. Recently, Yousaf and Bhatti
\cite{nyb12} observed more restricted unstable Newtonian and
post-Newtonian regimes in $f(R,T)$ gravity (as compared to $f(R)$
gravity) for the locally anisotropic collapsing stellar model.

The emergence of curvature singularity for the stellar systems has been discussed in
$f(R)$ gravity theory \cite{Arbuzova:2010iu, 6}. Houndjo \cite{7}
performed the cosmological reconstruction of $f(R,T)$ gravity
examining the transition of matter dominated epoch to the late time
accelerated regime. Alvarenga et al. \cite{8} formulated energy
conditions depending upon the attractive nature of gravity using
Raychaudhuri equation. Also, they investigated the viability of some
particular $f(R,T)$ gravity models using these energy conditions.
Azizi \cite{9} carried out the possibility whether static
spherically symmetric traversable wormhole geometries (which are
basically the exotic cosmic models) exist in $f(R,T)$ gravity.

In the study of accelerated universe expansion, the viscosity
effects due to matter configurations are quite vital and appearing
as the only non-adiabatic way in FRW models. Bulk viscosity disburse
negative pressure offering thus providing a platform for negative
pressure indicating repulsive gravity. During the particle creation
and formation of galaxies and clusters in the early universe,
neutrinos decouple from the cosmic fluid and viscosity arises in the
system \cite{10}. Naidu et al. \cite{11} studied the cosmological
model with FRW metric in the presence of viscosity in $f(R,T)$
gravity. Reddy et al. \cite{12} investigated the Kaluza-Klien
universe model in the presence of viscosity with the background of
$f(R,T)$ modified gravity theory. Sharif and his collaborators
\cite{13} have explored some physical processes with shear-free as
well as expansion and expansion-free self-gravitating collapsing
objects. Kiran and Reddy \cite{14} determined the solutions of field
equations in $f(R,T)$ gravity theory for Bianchi type III spatially
homogeneous model.

Nojiri and Odintsov \cite{z13} claimed that inflationary modified
high degrees of freedom quantities boosts up the evolution of
Schwarzschild-de Sitter black hole anti-evaporation in classical
background. Farinelli et al. \cite{z9} discussed equilibrium state
of the hydrostatic celestial objects and concluded that wide range
of compact objects exists in the nature of modified gravity. Guo et
al. \cite{z10} investigated dynamical behavior of spherical
relativistic collapse in modified gravity. Albareti et al.
\cite{z11} analyzed homogeneous cosmological models through
Raychaudhuri expressions and produce some viability constraints
coming due to modified gravity theory expansion regimes of the
universe. Hason and Oz \cite{z12} observed extended configurations
of Jeans instability condition for the relativistic systems for
normal and super fluids.

During the evolution of star model, a large amount of radiations
emit in the form of photons and neutrinos which gradually increases
as the evolution proceeds. The radiating energy can be characterized
in two approximations i.e., diffusion and free-streaming
approximation. The diffusion limit is applicable when the typical
length of the object is greater than the mean free path of the
particles responsible for the motion of energy. In that case, the
dissipation is described by a heat flow type vector while in the
other case it is characterized by an outflow of null fluid. Herrera
et al. \cite{15} investigated that the energy density should be
inhomogeneous if the system is involving with zero expansion
condition in non-dissipative fluid background. Herrera \cite{16}
explored some factors for a self-gravitating spherical star which
are important to describe the irregularities in the matter
distribution. Sharif and his collaborators \cite{17} have also
explored some factors describing the inhomogeneous density
distribution for self-gravitating objects with different matter
configurations.

In a recent paper, Yousaf et al. \cite{18} have formulated some
dynamical variables by splitting the Riemann tensor into its
constituent trace and trace-free scalar parts in $f(R,T)$ theory.
They have also discussed the evolution of shear and expansion using
the Raychaudhuri equation. This paper is organized in the following
manner. In the next section, we will provide some basic equations
including the action of this framework and equation of motion. In
section \textbf{3}, modified field equations, some kinematical and
dynamical quantities as well as modified Ellis equations are
formulated for the construction of our analysis in a systematic way.
Section \textbf{4} explores the irregularity factors with some
particular cases dissipative and non-dissipative matter
distribution. Finally, we conclude our results in the last section.

\section{$f(R,T)$ Gravity and Spherical Systems}

The notion of $f(R,T)$ gravity as a possible modifications in the
gravitational framework of GR received much attention of
researchers. This theory provides numerous interesting results in
the field of physics and cosmology like plausible explanation to the
accelerating cosmic expansion \cite{ya3, b2b, b2a}. The main theme
of this theory is to use an algebraic general function of Ricci as
well trace of energy momentum tensor in the standard EH action. It
can be written as \cite{3}
\begin{equation}\label{1}
S_{f(R,T)}=\int d^4x\sqrt{-g}[f(R,T)+L_M],
\end{equation}
where $g,~T$ are the traces of metric as well as standard GR
energy-momentum tensors, respectively while $R$ is the Ricci scalar.
There exists variety $L_M$ in literature which corresponds to
particular configurations of relativistic matter distributions.
Choosing $L_M=\mu$ (where $\mu$ is the system's energy density) and
making variation in the above equation with $g_{\alpha\beta}$, the
corresponding $f(R,T)$ field equations are given as follows
\begin{equation}\label{2}
{G}_{\alpha\beta}={{T}_{\alpha\beta}}^{\textrm{eff}},
\end{equation}
where
\begin{align*}\nonumber
{{T}_{\alpha\beta}}^{\textrm{eff}}&=\left[(1+f_T(R,T))T^{(m)}_{\alpha\beta}-\mu
g_{\alpha\beta}f_T(R,T)
-\left(\frac{f(R,T)}{R}-f_R(R,T)\right)\frac{R}{2}\right.+\\\nonumber
&\left.+\left({\nabla}_\alpha{\nabla}_
\beta+g_{\alpha\beta}{\Box}\right)f_R(R,T)\right]\frac{1}{f_R(R,T)}
\end{align*}
is a non-standard energy-momentum tensor representing modified
version of gravitational contribution coming from $f(R,T)$ extra
degrees of freedom while ${G}_{\alpha\beta}$ is an Einstein tensor.
Further, $\nabla_\alpha$ represents covariant derivation while
$f_T(R,T),~\Box$, $f_R(R,T)$ indicate
$\frac{df(R,T)}{dT},~\nabla_\alpha\nabla^\alpha$ and
$\frac{df(R,T)}{dR}$ operators, respectively.

We consider a spherical relativistic self-gravitating non-rotating
and non-static system. The metric of whose can be expressed with the
help of the following diagonal form form
\begin{equation}\label{3}
ds^2=-A^2(t,r)dt^{2}+B^2(t,r)dr^{2}+C^2d\theta^{2}
+C^2\sin^2\theta{d\phi^2}.
\end{equation}
It is assumed that this system is filled with shearing viscous,
locally anisotropic and radiating fluid. This fluid can be indicated
through the following configurations of the mathematical form
\begin{equation}\label{4}
T_{\alpha\beta}=P_{\bot}h_{\alpha\beta}+{\mu}V_\alpha V_\beta+\Pi
\chi_\alpha\chi_\beta+{\varepsilon}l_\alpha l_\beta+q(\chi_\beta
V_\alpha+\chi_\alpha V_\beta)-2{\eta}{\sigma}_{\alpha\beta},
\end{equation}
where $P_\bot$ is the tangential pressure, $\Pi\equiv P_r-P_\bot$,
$P_r$ is a fluid pressure along the radial direction. $\varepsilon$
is a radiation density, $q_{\beta}$ is a vector controlling heat
dissipation, $\sigma_{\alpha\beta}$ is a tensor controlling shearing
viscosity while $\eta$ is its coefficient. Further,
$h_{\alpha\beta}$ is the projection tensor defined as follows
\begin{eqnarray*}\nonumber
h_{\alpha\beta}=g_{\alpha\beta}+V_{\alpha}V_{\beta}
\end{eqnarray*}
The vectors $l^\gamma,~V^{\gamma}$ and $\chi^\gamma$ represent null
four-vector, fluid four-velocity and radial unit four-vector,
respectively. Under co-moving coordinates, these four-vectors can
evaluated as $V^{\gamma}=\frac{1}{A}\delta^{\gamma}_{0},~
\chi^{\gamma}=\frac{1}{C}\delta^{\gamma}_{1},~
l^\gamma=\frac{1}{A}\delta^{\gamma}_{0}+\frac{1}{B}\delta^{\gamma}_{1},~
q^\gamma=q(t,r)\chi^{\gamma}$. Moreover, they obey
\begin{eqnarray*}
&&V^{\beta}V_{\beta}=-1,\quad\chi^{\beta}\chi_{\beta}=1,
\quad\chi^{\beta}V_{\beta}=0,\\\nonumber &&V^\beta q_\beta=0, \quad
l^\beta V_\beta=-1, \quad l^\beta l_\beta=0.
\end{eqnarray*}
The scalar variable controlling expansion and contraction of matter
distribution is known as expansion scalar. This can be obtained
through $\Theta=V^{\alpha}_{~;\alpha}$ mathematical expression. For
Eq.(\ref{3}), it is found as follows
\begin{equation}\label{4a}
\Theta=\frac{1}{A}\left({\dot{B}}{B}^{-1}+{2\dot{C}}{C^{-1}}\right),
\quad
\sigma=\frac{-1}{A}\left({\dot{C}}{C}^{-1}-{\dot{B}}{B}^{-1}\right),
\end{equation}
where over dot notation stands for temporal partial derivation.

The $f(R,T)$ field equations (\ref{2}) for spherical non-static
interior (\ref{3}) are found as
\begin{align}\label{5}
G_{00}&=\frac{A^2}{f_{R}}\left[{\mu}+{\varepsilon}
-\frac{R}{2}\left(\frac{f}{R}-f_{R}\right)+\frac{\psi_{00}}{A^2}
\right],\\\label{6}
G_{01}&=\frac{AB}{f_{R}}\left[-(1+f_T)(q+{\varepsilon})
+\frac{\psi_{01}}{AB}\right],\\\label{7}
G_{11}&=\frac{B^2}{f_{R}}\left[\mu
f_T+(1+f_T)(P_r+\varepsilon-\frac{4}{3}\eta{\sigma})
+\frac{R}{2}\left(\frac{f}{R}-f_{R}\right)+\frac{\psi_{11}}{B^2}\right],\\\label{8}
G_{22}&=\frac{C^2}{f_{R}}\left[(1+f_T)({P_{\bot}}+\frac{2}{3}\eta{\sigma})+\mu
f_T
+\frac{R}{2}\left(\frac{f}{R}-f_{R}\right)+\frac{\psi_{22}}{C^2}\right],
\end{align}
where
\begin{align}\nonumber
\psi_{00}&=2\partial_{tt}f_R+\left(\frac{\dot{B}}{B}-2\frac{\dot{A}}{A}
+2\frac{\dot{C}}{C}\right)\partial_tf_R+\left(A^2\frac{B'}{B}-2AA'
-2A^2\frac{C'}{C}\right)\frac{\partial_rf_R}{B^2},\\\nonumber
\psi_{01}&=\partial_t\partial_rf_R-\frac{A'}{A}\partial_tf_R
-\frac{\dot{B}}{B}\partial_rf_R,\\\nonumber
\psi_{11}&=\partial_{rr}f_R-\frac{B^2}{A^2}\partial_{tt}f_R
+\left(B^2\frac{\dot{A}}{A}-2B^2\frac{\dot{C}}{C}-2B\dot{B}
\right)\frac{\partial_tf_R}{A^2} \\\nonumber
&+\left(\frac{A'}{A}+2\frac{C'}{C}
-2\frac{B'}{B}\right)\partial_rf_R,\\\nonumber
\psi_{22}&=-C^2\frac{\partial_{tt}f_R}{A^2}+\frac{C^2}{A^2}\left(
\frac{\dot{A}}{A}-3\frac{\dot{C}}{C}-\frac{\dot{B}}{B}\right)
\partial_tf_R+\frac{C^2}{B^2}\left(\frac{C'}{C}+\frac{A'}{A}
-\frac{B'}{B}\right)\partial_rf_R.
\end{align}
Here, the prime stands for radial partial differentiation.

Now, we are interested to evaluate expressions that would be helpful
to study the dynamical phases of spherical anisotropic radiating and
shearing viscous interiors in $f(R,T)$ gravity. It is seen that in
this gravitational theory, the divergence of energy momentum tensor
is non vanishing and is found to be
\begin{align}\label{9}
\nabla^{\alpha}T_{\alpha\beta}=\frac{f_T}{(1-f_T)}\left[
(\Theta_{\alpha\beta}+T_{\alpha\beta})\nabla^\alpha{\ln}f_T
-\frac{1}{2}g_{\alpha\beta}\nabla^\alpha{T}
+\nabla^\alpha\Theta_{\alpha\beta}\right].
\end{align}
The divergence of $f(R,T)$ energy-momentum tensor gives the
following couple of equations of motion
\begin{align}\nonumber
&\dot{\mu}\left(\frac{1+f_T+f_Rf_T}{f_R(1+f_T)}\right)-\frac{\mu}{f_R^2}\partial_tf_R
-\left\{\frac{f-Rf_R}{2}\right\}_{,0}+\left(\frac{\psi_{00}}{A^2}\right)_{,0}-\bar{q}'\frac{B}{A}\\\nonumber
&\times\frac{(1+f_T)}{f_R}-\frac{\bar{q}}{A^2}\left\{\frac{AB(1+f_T)}{f_R}\right\}_{,1}+\frac{1}{A^2}
\left(\frac{\psi_{01}}{f_R}\right)_{,1}-\frac{B\dot{B}}{A^2f_R}\left\{(1+f_T)\mu\right.\\\nonumber\
&\left.+\varepsilon+(1+f_T)\left(P_r+\varepsilon
+\frac{4}{3}\eta\sigma\right)+\frac{\psi_{00}}{A^2}+\frac{\psi_{11}}{B^2}\right\}
-\frac{2C\dot{C}}{A^2f_R}\left\{\bar{\mu}+\mu
f_T+(1\right.\\\nonumber\ &\left.+f_T)\left(P_\bot
+\frac{2}{3}\eta\sigma\right)+\frac{\psi_{00}}{A^2}+\frac{\psi_{22}}{C^2}\right\}+\left(\frac{B'}{B}
+\frac{A'}{A}-\frac{CC'}{B^2}+2\frac{AA'}{B^2}\right)\frac{1}{f_R}\\\label{10}
&\times\left\{(1+f_T)AB\bar{q}-\frac{\psi_{01}}{A^2}\right\}+\frac{1}{1+f_T}
\left\{(2\mu+\varepsilon)\partial_tf_T+\left(\dot{\mu}+\frac{\dot{T}}{2}\right)f_T\right\}=0,
\\\nonumber
&\frac{(1+f_T)}{f_R}\left\{\bar{P'_{r}}-\frac{4}{3}\eta\sigma'\right\}+\frac{\mu'f_T}{f_R}
+\frac{1}{(1+f_T)}\left(\bar{P_r}-\frac{4}{3}\eta\sigma-\mu\right)\partial_rf_T-\frac{A}{Bf_R}\bar{q}'\\\nonumber
&\times(1+f_T)-\left\{\frac{AB(1+f_T)}{f_R}\right\}_{,0}\frac{1}{B^2}+\frac{1}{B^2}
\left(\frac{\psi_{01}}{f_R}\right)_{,0}+\frac{\mu\partial_rf_T}{f_R}-\frac{\mu
f_T}{f_R^2}\partial_rf_R\\\nonumber
&+\frac{1}{f_R}\left(\bar{P_r}-\frac{4}{3}\eta\sigma\right)\left\{\partial_rf_T
-\frac{(1+f_T)}{f_R}\partial_rf_R\right\}+\left(\frac{f-Rf_R}{2}+\frac{\psi_{11}}{B^2}\right)_{,1}
-\frac{AA'}{B^2f_R}\\\nonumber &\times\left\{\mu
f_T+\bar{\mu}+(1+f_T)\left(\bar{P_r}-\frac{4}{3}\eta\sigma\right)+\frac{\psi_{11}}{B^2}
+\frac{\psi_{00}}{A^2}\right\}+\frac{2CC'}{B^2f_R}\left\{(1+f_T)\left(\bar{P_r}\right.\right.\\\nonumber
&\left.\left.-P_\bot-2\eta\sigma\right)+\frac{\psi_{11}}{B^2}
-\frac{\psi_{22}}{C^2}\right\}-\frac{f_T}{(1+f_T)}\left(\mu'+\frac{\partial_rT}{2}\right)
+\left(\frac{\dot{B}}{B}
+\frac{\dot{A}}{A}+\frac{2C\dot{C}}{A^2}\right.\\\label{11}
&+\left.2\frac{B\dot{B}}{A^2}\right)\left\{\frac{A}{B}\frac{(1+f_T)}{f_R}\bar{q}-\frac{\psi_{01}}{B^2f_R}\right\}=0.
\end{align}

The matter content within the spherical collapsing stellar geometry
can be defined through the general Misner-Sharp formula \cite{19}.
This is obtained as
\begin{equation}\label{12}
m=\left\{\frac{\dot{C}^2}{A^2}-\frac{C'^2}{B^2}+1\right\}\frac{C}{2},
\end{equation}
Before calculating its variations among adjacent surfaces of
spherical radiating fluid configurations, we first introduce some
useful operators. The operators corresponding to proper and radial
derivations are found as
\begin{eqnarray}\label{13}
D_{T}=\frac{1}{A} \frac{\partial}{\partial t},\quad
D_{C}=\frac{1}{C'}\frac{\partial}{\partial r}.
\end{eqnarray}
The relativistic velocity associated with the spherical stellar
structure can be found by using above mentioned proper derivative
operator. This turns out to be
\begin{eqnarray}\label{14}
U=D_{T}C=\frac{\dot{C}}{A}.
\end{eqnarray}
Now, we define $E$ as a ratio $\frac{C'}{B}$. From Eqs.(\ref{12})
and (\ref{14}), one can obtain
\begin{eqnarray}\label{15}
E=\sqrt{1+U^{2}-\frac{2m(t,r)}{C}}.
\end{eqnarray}
Using field equations and above two equations, the radial mass
variations is found as
\begin{align}\label{16}
D_Cm&=\frac{C^2}{2f_R}\left[\bar{\mu}-\frac{R}{2}\left(\frac{f}{R}-f_{R}\right)
+\frac{\psi_{00}}{A^2}+\frac{U}{E}\left\{(1+f_T)\bar{q}
-\frac{\psi_{01}}{AB}\right\}\right],
\end{align}
whose integration yields
\begin{align}\label{17}
m&=\frac{1}{2}\int^C_{0}\frac{C^2}{f_R}\left[\bar{\mu}-\frac{R}{2}
\left(\frac{f}{R}-f_{R}\right)
+\frac{\psi_{00}}{A^2}+\frac{U}{E}\left\{(1+f_T)\bar{q}
-\frac{\psi_{01}}{AB}\right\}\right]dC,
\end{align}
where over bar indicates the addition of radiation density in the
corresponding variable quantity. The particular combinations of
radiating matter parameters, $f(R,T)$ higher curvature terms and
energy density can be achieved via Misner-Sharp mass formulation.
This can be obtained, after using Eq.(\ref{17}), and is found as
follows
\begin{equation}\label{18}
\frac{3m}{C^3}=\frac{3}{2C^3}\int^r_{0}\left[\bar{\mu}
-\frac{R}{2}\left(\frac{f}{R}-f_{R}\right)
+\frac{\psi_{00}}{A^2}+\frac{U}{E}\left\{(1+f_T)\bar{q}
-\frac{\psi_{01}}{AB}\right\}C^2C'\right]dr.
\end{equation}
This equation can be recast to obtain a scalar related with the
tidal forces acting on the anisotropic radiating shearing viscous
spherical stellar system
\begin{align}\nonumber
\mathcal{E}&=\frac{1}{2f_R}\left[\bar{\mu}-(1+f_T)(\bar{\Pi}-2\eta\sigma)
-\frac{R}{2}\left(\frac{f}{R}-f_{R}\right)
+\frac{\psi_{00}}{A^2}-\frac{\psi_{11}}{B^2}
+\frac{\psi_{22}}{C^2}\right]\\\label{19} &-\frac{3m}{C^3},
\end{align}
where $\mathcal{E}$ is the Weyl scalar. It so happened that Weyl
scalar can be decomposed into its magnetic and electric
constituents. The magnetic part of the Weyl scalar is zero for
spherical matter distribution. However, electric part do exists. The
scalar, $\mathcal{E}$, is associated with this component of the Weyl
tensor. In this way, $\mathcal{E}$ describes the gravity effects
coming due to tidal forces in cosmos. Equation (\ref{19}) has
related tidal forces with the structural properties of the fluid
configurations and $f(R,T)$ extra curvature terms. Equation
(\ref{19}) has been evaluated by taking regular distribution of
fluid contents at the central point, i.e., $m(t,0)=0=C(t,0)$.

\section{Expansion-free Condition and $f(R,T)$ Ellis Equations}

In this section, we shall evaluate expansion-free constraint and
then discuss its meaning in the interpretation mysterious dark
universe. We then consider viable and well-consistent $f(R,T)$
gravity model. We then proceed forward our analysis by evaluating
well-known Ellis equations. The expansion-free equation can be
achieved by equating expansion scalar to zero. Thus, Eq.(\ref{4a})
yields
\begin{equation}\label{20}
\frac{\dot{B}}{B}=-\frac{2\dot{C}}{C},
\end{equation}
which upon integration gives
\begin{equation}\label{21}
B=\frac{h}{C^2},
\end{equation}
where $h$ is an arbitrary integration radial function.

Gravitational collapse is the phenomenon that takes place in this
accelerating expanding cosmos, when the state of hydrostatic
equilibrium of a celestial body is destroyed. If a stellar object is
massive enough such that the gas pressure is insufficient to support
it against gravitational forces then, the star undergoes
gravitational collapse giving birth to new stars. It is important to
stress that any self-gravitating stellar body would subject to
gravitational collapse once it bears through inhomogeneous and
irregular surface energy density. Therefore, in the collapse of
self-gravitating relativistic fluids, the role of energy density
inhomogeneity has gained much significance. If the fluid of
relativistic celestial interiors is expansion-free, then this study
may gain even more attention.

The expansion-free condition has produced several interesting
results at galactic and cosmological scales. Skripkin \cite{m8}
noticed the captivating process of cavity emergence within
non-radiating ideal relativistic matter field. It is seen from the
literature that in null expansion evolution, the innermost boundary
surface of the interior fluid configuration slides away from the
central point, thus conceiving vacuum Minkowskian core \cite{z1}.
The nullity of $\Theta$ is sufficient but not a necessary constraint
that guarantee the cavity emergence. The scenario of cavity
emergence have been explored in the literature \cite{z2} under some
kinematical constraints other than $\Theta=0$. Di Prisco et al.
\cite{z3} and Sharif and Yousaf \cite{z4} studied core formation
within the relativistic celestial locally anisotropic configurations
after its central explosion and demonstrated some expansion-free
relativistic solutions.

The possible implementations of null expansion condition is
anticipated for those astronomical settings where a Minkowskian core
is probably to be appear. In addition to this, during the process of
gravitational collapse, whenever the expansion-free matter moves
inside to reach the central point, there will be a strong shear
scalar blowup. Joshi et al. \cite{nya1} claimed that the apparent
horizon formation could be delayed due to the effects originated
from strong shear of the collapsing system. This suggests the
emergence of naked singularity (NS). Therefore, the study of
expansion-free relativistic interiors could provide an uncomplicated
platform for the analysis of NS appearance. NS is a spacetime
singularity that can be observed directly by a distant observer. It
represents the formation of extremely high curvature and strong
gravity regions and could provides a source of gravitational waves.
Are black holes (BHs) and NS observationally distinguished from each
other? In this perspective, Virbhadra et al. \cite{nya2} gave a very
useful mathematical tool to understand the NS physics.

Virbhadra and Ellis \cite{nya3} established that one can
observationally differentiate NS from BHs by analyzing the
corresponding characteristics of gravitational lensing (GL). Claudel
et al. \cite{nya4} demonstrated that any photon relativistic
spherical body could be around the BH only if it obeys a reasonable
energy condition. For the observational study of cosmic censorship
hypothesis, GL could provide a reliable direction. In this context,
it is seen from \cite{nya5} that BH and NS of the same symmetry and
Arnowitt-Deser-Misner mass yield variety of different images of the
same source of light. Further, time of image delays because of GL by
a BH is greater than that of NS. This asserts that one can get
smaller time delays by choosing extreme values of nakedness
variables. NS could also provide images with negative delays of time
\cite{nya6}.

Due to zero expansion, matter sources could be effective for the
voids explanation. Voids are, so called, underdense regions
incorporating substantial amount of information on the cosmological
environment \cite{v1}. Voids offers a reliable guide to discuss the
cosmic structure appearance at large scales. In comparison with GR,
they are more rich in modified gravity \cite{v2}. Wiltshire
\cite{z5} claimed that the actual picture of cosmos constitutes
spongelike structural bodies in which voids have a dominant role.
Further, some cosmological indications asserts that about $40-50$\%
volume of the today cosmos is endowed with cosmological voids with a
scale $30\textmd{h}^{-1}\textmd{Mpc}$, where h is the
non-dimensional Hubble parameter,
$H_0=100\textmd{h~km~sec}^{-1}\textmd{Mpc}^{-1}$.

The evident relevance of such mentioned effects could reinforces the
interest of the problem mentioned in this paper.

The fascinating phenomenon of accelerating cosmic expansion could be
described by taking into account extended gravity models involving
curvature matter coupling, like $f(R,T)$ gravity. For theoretically
and cosmologically consistent $f(R,T)$ gravity, the choice of
$f(R,T)$ function is very crucial. We are considering the following
particular $f(R,T)$ model form
\begin{equation}\label{22}
f(R,T)=f_1(R)+f_2(T),
\end{equation}
This model form does involve direct minimal curvature matter
coupling. This could be assumed as a possible correction in the
well-known $f(R)$ gravity. Here, we take a linear choice of $f_2$
due to which some striking outcomes can be observed on the basis of
non-trivial coupling as compared to $f(R)$ gravity. Thus, we assume
$f_2(T)=\nu T$, where $\nu$ is a constant. The Lagrangian with this
background of $f_2$ has broadly been examined by many relativistic
astrophysicists. Harko et al. \cite{z6} obtained some cosmic
solutions depicting clear accelerating expanding picture of the
universe framed within $f_2=\nu T$.

Now, we are interested to take a physical feasible generic Ricci
invariant function. These may give birth to the existence of some
new spherical models. A cosmological viable model needs to obey the
big-bang nucleosynthesis, radiation as well as matter dominated
regimes. Also, they should expect to allow cosmological
perturbations consistent with cosmic restrictions emerging from
anisotropies in cosmic microwave background. In this realm, we
consider power law Ricci scalar corrections, i.e., $f(R)=R+\lambda
R^n$, where $\lambda\in \mathbb{R}^+$ with $\mathbb{R}^+$ is the set
of positive real numbers and $n$ is a constant. Depending upon the
selection of $n$, this model has some physical descriptions. For
instance, this model, for $n=2$, could depicts exponential behavior
of the early cosmic expansion as proposed by Starobinsky
\cite{staro}. Such $f(R)$ model could draw dark matter (for
$\lambda=\frac{1}{6M^2}$ \cite{v40} with
$M=2.7\times10^{-12}\mathrm{GeV}$ \cite{v41}) and DE effects in the
gravitational theory. Furthermore, gravity induced under $n=2$
\cite{z8} and $n=3$ \cite{asta1} support the existence of more
massive compact objects as compared to GR. The $f(R)$ tanh
corrections have also been investigated in the study of stellar
collapse \cite{zbb1}. However, the negative $n$ values could helps
to explain dynamics of stellar object in the presence of late time
accelerating cosmic expansion corrections \cite{ya3}.

In order to delve with the survival of the regular energy density
over the dissipative spherical celestial object, we now calculate
couple of well-known equation by following the procedure introduced
by Ellis \cite{z7}. These expressions in the background of dark
source $f(R,T)$ corrections can be found by using
Eqs.(\ref{5})-(\ref{8}), (\ref{12}), (\ref{13}), (\ref{19}) and
(\ref{22}) as
\begin{align}\nonumber
&\left[\mathcal{E}-\frac{1}{2(1+n\lambda
{R}^{n-1})}\left\{\bar{\mu}-(1+\nu)(\bar{\Pi}-2\eta\sigma)
-\frac{(1-n)}{2}\lambda
{R}^n-\frac{\nu}{2}{T}+\frac{\varphi_{00}}{A^2}\right.\right.\\\nonumber
&-\left.\left.\frac{\varphi_{11}}{B^2}+\frac{\varphi_{22}}{C^2}
\right\}\right]_{,0}=\frac{3\dot{C}}{C}\left[\frac{1}{2(1+n\lambda
{R}^{n-1})}\left\{\bar{\mu}+(1+\nu)\left(P_\bot-\frac{2}{3}\eta\sigma\right)
+\mu\nu\right.\right.\\\label{23}
&+\left.\left.\frac{\varphi_{00}}{A^2}+\frac{\varphi_{22}}{C^2}\right\}-\mathcal{E}\right]+\frac{3AC'}{2BC(1+n\lambda
{R}^{n-1})}\left\{(1+\nu)\bar{q}-\frac{\varphi_{01}}{AB}\right\},\\\nonumber
&\left[\mathcal{E}-\frac{1}{2(1+n\lambda
{R}^{n-1})}\left\{\bar{\mu}-(1+\nu)(\bar{\Pi}-2\eta\sigma)
-\frac{(1-n)}{2}\lambda
{R}^n-\frac{\nu}{2}{T}+\frac{\varphi_{00}}{A^2} \right.\right.
\\\nonumber
&-\left.\left.\frac{\varphi_{11}}{B^2}+\frac{\varphi_{22}}{C^2}\right\}\right]'
=-\frac{3C'}{C}\left[\mathcal{E}+\frac{1}{2(1+n\lambda
{R}^{n-1})}\left\{(1+\nu)(\bar{\Pi}-2\eta\sigma)+\frac{\varphi_{11}}{B^2}
\right.\right.\\\label{24}
&-\left.\left.\frac{\varphi_{22}}{C^2}\right\}\right]-\frac{3B\dot{C}}{2AC(1+n\lambda
{R}^{n-1})}\left\{(1+\nu)\bar{q}-\frac{\varphi_{01}}{AB}\right\},
\end{align}
where $\varphi_{ii}$ encapsulate $f(R,T)$ extra degrees of freedom
involved in the evolution of shearing viscous radiating spherical
body. These quantities can be evaluated by considering
Eqs.(\ref{5})--(\ref{8}) and (\ref{22}) accordingly.

\section{Irregularities in the Dynamical System}

In this section, we shall calculate some irregularity factors that
causes the appearance of irregularities over the surface of stellar
spherical system with $f(R,T)$ background. The system enters in the
collapsing window once celestial surface suffers energy density
inhomogeneities. Therefore, the understanding of the system's
collapsing nature is directly related to the exploration of
irregularity factors. For this purpose, we assume that our stellar
spherical relativistic system is in complete homogenous phase. We
shall take some specific choices of matter fields framed within dark
source terms coming from $f(R,T)$ model. As $f(R,T)$ field equations
are highly non-linear, therefore we would confine ourself at the
constant values of trace of stress-energy tensor as well as
cosmological Ricci scalar. These are represented by putting over
tilde over the respective quantities. We shall also calculate
irregularity factors for those spherical relativistic interior that
continue their evolutions by establishing central Minkowskian
cavity. This would be achieved by taking expansion-free condition in
the corresponding equations. We shall classify our investigation in
two scenarios, i.e., dissipative.radiating and
non-dissipative/non-radiating systems as follows:

\subsection{Non-radiating Matter}

Here, we deal with adiabatic non-interacting, ideal and locally
anisotropic forms of relativistic matter distributions coupled
framed within $f(R,T)$ background.

\subsubsection{Non-interacting Relativistic Particles}

This subsection addresses geodesically moving non-interacting fluid
configurations. So, we take all pressure gradients effects to be
zero $\hat{P}_r=0=P_\bot=\hat{q}$ along with $A=1$. Then $f(R,T)$
Ellis equations (\ref{23}) and (\ref{24}) reduce to
\begin{align}\nonumber
&\left[\mathcal{E}-\frac{1}{2(1+n\lambda \tilde{R}^{n-1})}\left\{\mu
-\frac{(1-n)}{2}\lambda\tilde{R}^n-\frac{\nu}{2}\tilde{T}\right\}\right]_{,0}=\frac{3\dot{C}}{C}
\left[\frac{1}{2(1+n\lambda \tilde{R}^{n-1})}\right.\\\nonumber
&\left.\left\{\mu
(1+\nu)\right\}-\mathcal{E}\right],\\\nonumber
&\left[\mathcal{E}-\frac{1}{2(1+n\lambda \tilde{R}^{n-1})}\left\{\mu
-\frac{(1-n)}{2}\lambda\tilde{R}^n-\frac{\nu}{2}\tilde{T}\right\}\right]'=-3\frac{C'}{C}\mathcal{E}
\end{align}
Using Eqs.(\ref{4a}), (\ref{10}) and (\ref{22}) in above equations,
we have
\begin{align}\label{25}
&\dot{\mathcal{E}}+\frac{3\dot{C}}{C}\mathcal{E}=\frac{\mu(1+\lambda)}{2(1+n\lambda
\tilde{R}^{n-1})}\left[\frac{3\dot{C}}{C}+\frac{(1+\nu)B^2C^2}{\{1+(1+n\lambda
\tilde{R}^{n-1})(1+\nu)\}}\Theta\right],\\\label{26}
&{\mathcal{E}'}+\frac{3{C'}}{C}\mathcal{E}=\frac{\mu'}{2(1+n\lambda
\tilde{R}^{n-1})}.
\end{align}
It can seen from Eq.(\ref{26}) that, if $\mu'=\mu(t)$ then
\begin{align}\nonumber
&{\mathcal{E}}=0,
\end{align}
thereby indicating that existence of Weyl scalar is directly
proportional to the existence of regular energy density of
non-interacting self-gravitating particles. This is the very result
as found in GR by many relativistic astrophysicists. Thus, we
conclude that $f(R,T)$ extra curvature terms has not altered or
disturb the Weyl curvature role in the conformally flat solutions of
dust relativistic cloud. Now, we solve Eq.(\ref{25}) to investigate
that which quantities are infact making impact over the contribution
of Weyl scalar in $f(R,T)$ gravity. The solution of Eq.(\ref{25})
yields
\begin{align}\label{27}
{\mathcal{E}}&=\frac{(1+\nu)}{2(1+n\lambda
\tilde{R}^{n-1})C^3}\int_0^t\left[3C^2\dot{C}+\frac{(1+\nu)B^2C^5}{\{1+(1+n\lambda
\tilde{R}^{n-1})(1+\nu)\}}\Theta\right]\mu dt.
\end{align}
This shows that Weyl scalar for dust particles in $f(R,T)$ model is
directly related with the temporal integrals of energy density and
expansion scalar. If we take null expansion scenario, then above
equation gives
\begin{align}\label{28}
{\mathcal{E}}&=\frac{3(1+\nu)}{2(1+n\lambda
\tilde{R}^{n-1})C^3}\int_0^t\mu C^2\dot{C} dt.
\end{align}
The relativistic systems that are evolving by encapsulating
Minkowskian core in the universe should satisfy the above constraint
in order to enter in the inhomogeneous phase. In other words, for
the regular distribution of dust expansion-free particles in
$f(R,T)$ gravity, one needs to take Eq.(\ref{28}) to be zero.

\subsubsection{Isotropic Fluid}

Here, we consider the case of ideal self-gravitating fluid in the
environment of $f(R,T)$ gravity. The extended Ellis equations
(\ref{23}) and (\ref{24}) give rise to the following set of
differential equations
\begin{align}\nonumber
&\left[\mathcal{E}-\frac{1}{2(1+n\lambda \tilde{R}^{n-1})}\left\{\mu
-\frac{(1-n)}{2}\lambda\tilde{R}^n-\frac{\nu}{2}\tilde{T}\right\}\right]_{,0}=\frac{3\dot{C}}{C}
\left[\frac{1}{2(1+n\lambda \tilde{R}^{n-1})}\right.\\\nonumber
&\left.\left\{(\mu+P)(1+\nu)\right\}-\mathcal{E}\right],\\\nonumber
&\left[\mathcal{E}-\frac{1}{2(1+n\lambda \tilde{R}^{n-1})}\left\{\mu
-\frac{(1-n)}{2}\lambda\tilde{R}^n-\frac{\nu}{2}\tilde{T}\right\}\right]'=-3\frac{C'}{C}\mathcal{E}.
\end{align}
Equations (\ref{4a}) and (\ref{11}) provide
\begin{align}\label{29}
&\dot{\mathcal{E}}+\frac{3\dot{C}}{C}\mathcal{E}=\frac{(1+\lambda)(\mu+P)}{2(1+n\lambda
\tilde{R}^{n-1})}\left[\frac{3\dot{C}}{C}+\frac{(1+\nu)B^2C^2}{\{1+(1+n\lambda
\tilde{R}^{n-1})(1+\nu)\}}\Theta\right],\\\label{30}
&{\mathcal{E}'}+\frac{3{C'}}{C}\mathcal{E}=\frac{\mu'}{2(1+n\lambda
\tilde{R}^{n-1})}.
\end{align}
It is seen from the second of above equation that energy density
will be regular as long as $\mathcal{E}=0$. However, the solution of
Eq.(\ref{29}) yields
\begin{align}\label{31}
{\mathcal{E}}&=\frac{(1+\nu)}{2(1+n\lambda
\tilde{R}^{n-1})C^3}\int_0^t\left[3C^2\dot{C}+\frac{(1+\nu)B^2C^5}{\{1+(1+n\lambda
\tilde{R}^{n-1})(1+\nu)\}}\Theta\right](\mu+P)dt.
\end{align}
This indicates that the influence of tidal forces is controlled by
the linear combination of system energy density and locally
isotropic pressure gradient. This also highlights the importance of
expansion scalar in the modeling of homogeneous spherical geometry
coupled with isotropic matter configurations in the presence of
$f(R,T)$ corrections. However, if we eliminate this scalar with the
help of Eq.(\ref{20}), then we have
\begin{align}\label{32}
{\mathcal{E}}&=\frac{3(1+\nu)}{2(1+n\lambda
\tilde{R}^{n-1})C^3}\int_0^t(\mu+P) C^2\dot{C} dt.
\end{align}
This suggests that pressure gradient has increased the impact of
tidal forces over the isotropic spherical stellar interior. Further,
$f(R,T)$ corrections tends to reduce the influence of Weyl scalar
due to its non-attractive nature.

\subsubsection{Anisotropic Fluid}

This subsection is aimed to extend our previous work. Here, we
introduce effects of anisotropic stresses, thus $\Pi\neq0$ in our
analysis. In this realm, $f(R,T)$ Ellis equations (\ref{23}) and
(\ref{24}) take the following forms
\begin{align}\nonumber
&\left[\mathcal{E}-\frac{1}{2(1+n\lambda \tilde{R}^{n-1})}\left\{\mu
-(1+\nu)\Pi-\frac{(1-n)}{2}\lambda\tilde{R}^n-\frac{\nu}{2}\tilde{T}\right\}\right]_{,0}=\frac{3\dot{C}}{C}
\\\nonumber
&\times\left[\frac{1}{2(1+n\lambda
\tilde{R}^{n-1})}\left\{(\mu+P_\bot)(1+\nu)\right\}-\mathcal{E}\right],\\\nonumber
&\left[\mathcal{E}-\frac{1}{2(1+n\lambda \tilde{R}^{n-1})}\left\{\mu
-(1+\nu)\Pi-\frac{(1-n)}{2}\lambda\tilde{R}^n-\frac{\nu}{2}\tilde{T}\right\}\right]'=-3\frac{C'}{C}
\\\nonumber
&\times\left[\mathcal{E}+\frac{(1+\nu)\Pi}{2(1+n\lambda
\tilde{R}^{n-1})}\right],
\end{align}
which can be manipulated, after using Eq.(\ref{11}), in the
following forms
\begin{align}\nonumber
&\left[\mathcal{E}+\frac{(1+\lambda){\Pi}}{2(1+n\lambda
\tilde{R}^{n-1})}
\right]_{,0}+3\left[\mathcal{E}+\frac{(1+\lambda){\Pi}}{2(1+n\lambda
\tilde{R}^{n-1})}
\right]\frac{\dot{C}}{C}=\frac{3[\mu+(1+\nu)P_r]\dot{C}}{2C(1+n\lambda
\tilde{R}^{n-1})}\\\label{33} &+\frac{(1+\nu)^2(1+n\lambda
\tilde{R}^{n-1})^{-1}}{2A\{1+(1+n\lambda
\tilde{R}^{n-1})(1+\nu)\}}\left[(\mu+P_r)B^2C^5\Theta+\frac{2C\Pi\dot{C}}{A}\right],\\\label{34}
&\left[\mathcal{E}+\frac{(1+\lambda){\Pi}}{2(1+n\lambda
\tilde{R}^{n-1})}
\right]_{,1}+3\left[\mathcal{E}+\frac{(1+\lambda){\Pi}}{2(1+n\lambda
\tilde{R}^{n-1})} \right]\frac{{C'}}{C}=\frac{\mu'}{2(1+n\lambda
\tilde{R}^{n-1})}.
\end{align}
It is well-known from the working of several relativistic
astrophysicists that in GR \cite{m17} as well in $f(R,T)$ \cite{18},
one can break Riemann tensor into couple of tensors, namely,
$X_{\alpha\beta}$ and $Y_{\alpha\beta}$. The trace-less part of
$X_{\alpha\beta}$ yields the following following (for details please
see \cite{18})
\begin{align}\label{35}
X_{TF}=-\mathcal{E}-\frac{(1+\lambda){\Pi}}{2(1+n\lambda
\tilde{R}^{n-1})}.
\end{align}
It is seen that some terms involved in Eqs.(\ref{33}) and (\ref{34})
has the same configurations as that of the trace-less part of the
2nd dual of Riemann curvature tensor mentioned in Eq.(\ref{32}). In
this context, Eqs.(\ref{33}) and (\ref{34}) can be recasted as
\begin{align}\nonumber
\dot{X}_{TF}+\frac{3X_{TF}\dot{C}}{C}&=-\frac{3[\mu+(1+\nu)P_r]\dot{C}}{2C(1+n\lambda
\tilde{R}^{n-1})}-\frac{(1+\nu)^2(1+n\lambda
\tilde{R}^{n-1})^{-1}}{2A\{1+(1+n\lambda
\tilde{R}^{n-1})(1+\nu)\}}\\\label{36}
&\times\left[(\mu+P_r)B^2C^5\Theta+\frac{2C\Pi\dot{C}}{A}\right],\\\label{37}
X'_{TF}+\frac{3X_{TF}{C'}}{C}&=\frac{-\mu'}{2(1+n\lambda
\tilde{R}^{n-1})}.
\end{align}
The second of above equation points out if $\mu'=0$ the $X_{TF}=0$
and vice versa. This indicates $X_{TF}$ as an entity supervising
inhomogeneities in the energy density of the anisotropic spherical
fluids. This results supports the consequences of \cite{18}. Now, we
are interested to find out that to which factors this $X_{TF}$
further depends, in the presence of dark source terms due to
$f(R,T)$ gravity. The solution of Eq.(\ref{36}) yields
\begin{align}\nonumber
X_{TF}&=\frac{-3}{2(1+n\lambda
\tilde{R}^{n-1})C^3}\int_0^t[\mu+(1+\nu)P_r]C^2\dot{C}dt \\ \nonumber
&-\frac{(1+\nu)^2(1+n\lambda
\tilde{R}^{n-1})^{-1}}{2A\{1+(1+n\lambda
\tilde{R}^{n-1})(1+\nu)\}} \\\label{38}
&\times\int_0^t\left[(\mu+P_r)B^2C^5\Theta+\frac{2C\Pi\dot{C}}{A}\right]dt.
\end{align}
This points out the importance of pressure anisotropy and expansion
scalar in the modeling of regular energy density of the celestial
spherical geometry in $f(R,T)$ gravity. Now, using Eq.(\ref{20}) and
Eq.(\ref{21}), we get
\begin{align}\nonumber
X_{TF}&=\frac{-3}{2(1+n\lambda
\tilde{R}^{n-1})C^3}\int_0^t[\mu+(1+\nu)P_r]C^2\dot{C}dt
\\\label{39} &-\frac{(1+\nu)^2(1+n\lambda
\tilde{R}^{n-1})^{-1}}{A^2\{1+(1+n\lambda \tilde{R}^{n-1})(1+\nu)\}}
\int_0^tC\Pi\dot{C}dt.
\end{align}
This provides that inhomogeneity factor, i.e., $X_{TF}$ depends upon
anisotropic pressure gradients in the scenario of $f(R,T)$ gravity.
Since we know that in the null expansion stellar body, the central
point is covered by another metric appropriately joined with the
rest of the matter distributions.

\subsection{Radiating Shearing Viscous Non-Interacting Particles}

This subsection discusses the irregularity factors in the realm of
dissipation with both free streaming and diffusion limits, but with
a special case of viscous particles. Therefore, we consider
$P_r=0=P_\bot$ in the matter field and the evolution is
characterized by geodesics. This assumption is well established in
the background of some theoretical developments. Then,
Eqs.(\ref{23}) and (\ref{24}) give
\begin{align}\nonumber
&\left[\mathcal{E}-\frac{1}{2(1+n\lambda \tilde{R}^{n-1})}\left\{\mu
-2(1+\nu)\eta\sigma-\frac{(1-n)}{2}\lambda\tilde{R}^n-\frac{\nu}{2}\tilde{T}\right\}\right]_{,0}=\frac{3\dot{C}}{C}
\\\label{40a}
&\times\left[\frac{1}{2(1+n\lambda
\tilde{R}^{n-1})}\left\{\bar{\mu}-\frac{2}{3}(1+\nu)\eta\sigma+\mu\nu\right\}-\mathcal{E}\right]
+\frac{3A(1+\nu)\bar{q}C'}{2BC(1+n\lambda
\tilde{R}^{n-1})},\\\nonumber
&\left[\mathcal{E}-\frac{1}{2(1+n\lambda
\tilde{R}^{n-1})}\left\{\bar{\mu}
+2(1+\nu)\eta\sigma-\frac{(1-n)}{2}\lambda\tilde{R}^n-\frac{\nu}{2}\tilde{T}\right\}\right]'=-3\frac{C'}{C}
\\\label{40b}
&\times\left\{\mathcal{E}+\frac{(1+\nu)\eta\sigma}{(1+n\lambda
\tilde{R}^{n-1})}\right\}-\frac{3B(1+\nu)\bar{q}\dot{C}}{2AC(1+n\lambda
\tilde{R}^{n-1})}.
\end{align}
It has been investigated from the above equation that the quantity
which is controlling irregularities is $\Psi$ defined as follows
\begin{align}\label{40}
&\Psi\equiv\mathcal{E}+\frac{(1+\nu)}{C^3(1+n\lambda
\tilde{R}^{n-1})}\left[\eta\int_0^r\left(\sigma'-\frac{3C'}{C}\sigma\right)C^3dr
-\frac{3}{2}\int_0^rB\dot{C}\bar{q}C^2dr\right]
\end{align}
Thus, If we there is a regular configurations of energy density
i.e., $\mu'=\mu(t)$, then $\Psi=0$, and vice versa. Thus in order to
enter in the homogeneous window by the radiating dust cloud, it
should vanish the above quantity $\Psi$. It can be seen that $\Psi$
is controlled shearing viscosity and heat flux. Making use of
Eqs.(\ref{4a}) and (\ref{11}) in Eq.(\ref{40a}), the $\Psi$
evolution equation is found as follows
\begin{align}\nonumber
&\dot{\Psi}-\frac{\dot{\Omega}}{C^3}=\frac{(1+\nu)^2(1+n\lambda
\tilde{R}^{n-1})^{-1}}{2\{1+(1+n\lambda
\tilde{R}^{n-1})(1+\nu)\}}\left[\bar{q}'B+B^2C^2\left(\bar{\mu}+\frac{2}{3}\eta\sigma\right)\Theta
+B\right.\\\nonumber
&\times\left.\left(\varepsilon+\frac{2}{3}\eta\sigma\right)\dot{B}\right]
+\frac{\dot{\varepsilon}}{2(1+n\lambda
\tilde{R}^{n-1})}\left\{1-\frac{(1+\nu)}{\{1+(1+n\lambda
\tilde{R}^{n-1})(1+\nu)\}}\right\}\\\nonumber
&+\frac{\eta(1+\nu)}{(1+n\lambda
\tilde{R}^{n-1})}\left(\dot{\sigma}+\frac{\dot{C}}{C}\sigma\right)
+\frac{3\dot{C}}{2C(1+n\lambda
\tilde{R}^{n-1})}\{\varepsilon+\mu(1+\nu)\}-\frac{3\dot{C}}{C}\Psi\\\label{41}
&+\frac{(1+\nu)\bar{q}}{2(1+n\lambda
\tilde{R}^{n-1})}\left\{\frac{3C'}{C}-\frac{(1+\nu)}{\{1+(1+n\lambda
\tilde{R}^{n-1})(1+\nu)\}}\left(\frac{B'}{B}-\frac{CC'}{B^2}\right)\right\},
\end{align}
whose solution leads to
\begin{align}\nonumber
&{\Psi}=\frac{1}{C^3}\int_0^t\left[\dot{\Omega}+\left\{\frac{(1+\nu)^2(1+n\lambda
\tilde{R}^{n-1})^{-1}C^3}{2\{1+(1+n\lambda
\tilde{R}^{n-1})(1+\nu)\}}\left\{B^2C^2\left(\bar{\mu}+\frac{2}{3}\eta\sigma\right)\Theta+\bar{q}'B
\right.\right.\right.\\\nonumber
&+\left.\left.\left.\left(\varepsilon+\frac{2}{3}\eta\sigma\right)B\dot{B}\right\}\right\}
+\frac{\dot{\varepsilon}C^3}{2(1+n\lambda
\tilde{R}^{n-1})}\left\{1-\frac{(1+\nu)}{\{1+(1+n\lambda
\tilde{R}^{n-1})(1+\nu)\}}\right\}\right.\\\nonumber
&+\left.\frac{\eta(1+\nu)}{(1+n\lambda
\tilde{R}^{n-1})}\left(\dot{\sigma}+\frac{\dot{C}}{C}\sigma\right)C^3
+\frac{3\dot{C}C^2}{2(1+n\lambda
\tilde{R}^{n-1})}\{\varepsilon+\mu(1+\nu)\}\right.\\\label{42}
&+\left.\frac{(1+\nu)\bar{q}}{2C^3(1+n\lambda
\tilde{R}^{n-1})}\left\{\frac{3C'}{C}-\frac{(1+\nu)}{\{1+(1+n\lambda
\tilde{R}^{n-1})(1+\nu)\}}\left(\frac{B'}{B}-\frac{CC'}{B^2}\right)\right\}\right].
\end{align}
For expansion-free condition, the inhomogeneity factor for the
viscous dissipative system is found as follows
\begin{align}\nonumber
&{\Psi}=\frac{1}{C^3}\int_0^t\left[\dot{\Omega}+\left\{\frac{(1+\nu)^2(1+n\lambda
\tilde{R}^{n-1})^{-1}C^3}{2\{1+(1+n\lambda
\tilde{R}^{n-1})(1+\nu)\}}\left\{\frac{h\bar{q}'}{C^2}-\frac{2h^2\dot{C}}{C^5}
\left(\varepsilon+\frac{2}{3}\eta\sigma\right)\right\}\right\}\right.\\\nonumber
&+\left. +\frac{\dot{\varepsilon}C^3}{2(1+n\lambda
\tilde{R}^{n-1})}\left\{1-\frac{(1+\nu)}{\{1+(1+n\lambda
\tilde{R}^{n-1})(1+\nu)\}}\right\}+\frac{\eta(1+\nu)}{(1+n\lambda
\tilde{R}^{n-1})}\right.\\\nonumber
&\times\left.\left(\dot{\sigma}+\frac{\dot{C}}{C}\sigma\right)C^3
+\frac{3\dot{C}C^2}{2(1+n\lambda
\tilde{R}^{n-1})}\{\varepsilon+\mu(1+\nu)\}+\frac{(1+\nu)\bar{q}}{2C^3(1+n\lambda
\tilde{R}^{n-1})}\left\{\frac{3C'}{C}\right.\right.\\\label{43}
&\left.\left.-\frac{(1+\nu)}{\{1+(1+n\lambda
\tilde{R}^{n-1})(1+\nu)\}}\left(\frac{h'}{h}-\frac{2C'}{C}-\frac{C^5C'}{h^2}\right)\right\}\right].
\end{align}
This asserts the importance of matter parameters as the irregularity
factor has some correspondence with the fluid variables, especially
shearing viscosity, heat flux, as well as structural properties of
the system.

\section{Conclusions}

In this paper, we have studied the impact of modified gravity on the
distribution of matter configuration for a self-gravitating
spherical star. The disturbance in the hydrostatic equilibrium of a
celestial object leads to homogeneous or inhomogeneous matter state.
We have taken into consideration the spherically symmetric source in
the gravitational field of $f(R,T)$ gravity. The geometry is filled
with imperfect fluid due to anisotropic stresses and dissipation
which is designed in both limits i.e., diffusion and free-streaming
limit. We have constructed the modified field equations and
corresponding dynamical equations using conservation laws. Some
kinematical and dynamical quantities are formulated to explain the
evolutionary development of such objects. The mass function using
the Misner-Sharp \cite{19} approach is calculated for our spherical
object and the curvature tensor as well as the Weyl tensor are
explored in this framework. It is found that the Weyl tensor have
its constituent tensor like its electric and magnetic parts. Its
magnetic part vanishes due to symmetry of the under considered
problem while only its electric part exist.

It is well-established fact that the Weyl tensor is responsible for
the emergence of tidal forces which makes the object to be more
inhomogeneous with the evolution of time. In our case, the spherical
system suffers the inhomogeneous states due the presence of its
electric part only. We have established an explicit expression
between the Weyl tensor and the matter variables like heat flux and
anisotropic stresses etc., which holds a significant importance in
the light of Penrose's proposal \cite{20}. Penrose provided the idea
of relationship between the Weyl tensor with inhomogeneities in
energy density and isotropic pressure, however, such a link is no
longer valid in the presence of anisotropies. In this manuscript, we
have established such a relation between the Weyl tensor and fluid
parameters in the background of higher order curvature terms
emerging due to $f(R,T)$ gravitational field.

We have also disintegrated the curvature tensor into its constituent
parts using the comoving coordinates. These are found to be
structure scalars as already obtained in the framework of $f(R)$
gravity. These scalars have gained significant importance in the
light of Newtonian and general relativistic star models. It is also
observed that these scalars are used to find the solutions of the
Einstein field equations \cite{21}. Moreover, these scalars are also
used to discuss the irregular distribution of matter density. It is
still unclear that how different physical factors emerging in fluid
configuration can affect the production of inhomogeneities in energy
density. Here, we have found some factors creating the
irregularities in the matter distribution with $f(R,T)$ extra
curvature ingredients. Our analysis will strictly depend upon two
differential equations emerging from the explicit expression of Weyl
tensor with matter variables and the mass function. These equations
are carried out by using the Ellis's procedure as adopted in his
paper. We have constructed our analysis to demonstrate the
inhomogeneities in two regimes i.e., with dissipative and
non-dissipative fluid. The results obtained in the particular cases
of dust, isotropic and anisotropic matter are given as follows:

(i)
In the absence of dissipative effects and with dust cloud and $f(R,T)$ dark source terms, we found that the evolutionary motion of the celestial bodies will be homogeneous if the Weyl scalar is zero with extra curvature terms of the theory. In other words, the Weyl tensor due to its electric part and the impact of modified gravity makes the system more inhomogeneous in the gravitational arrow of time. This result can be seen from Eqs.(\ref{27}) and (\ref{28}), i.e., if we have a homogeneous matter distribution then Weyl tensor and dark source term should vanish.

(ii)
By increasing the complexity in the previous case with isotropic pressure, we observed the same factors creating the irregularities in the density distribution but in the presence of isotropic pressure.

(iii)
For non-dissipative anisotropic fluid model, it is found that a linear combination of matter profile is now responsible for the emergence of density inhomogeneity. Further, we have examined that such linear combination corresponds to one of the structure scalars as obtained in Eq.(\ref{39}).

(iv)
For dissipative dust cloud case, we have factor controlling the density distribution as a combination of geometrical and physical variables in the background of $f(R,T)$ gravity theory as obtained in Eq.(\ref{42}) and (\ref{43}).

We mention that this study can be generalize to study the
density inhomogeneity to $f(R,T,R_{\mu\nu}T^{\mu\nu})$ gravity. All
of our results reduce to general relativity if we take $f(R)=R$.

\section*{Acknowledgments}

We would like to sincerely appreciate the kind encouragements on
this work of Professor Muhammad Sharif. This work was partially
supported by the JSPS Grant-in-Aid for Young Scientists (B) \#
25800136 and the research-funds presented by Fukushima University
(K.B.).

\end{document}